\documentclass{article}
\usepackage{spconf,amsmath,amssymb,graphicx,hyperref,dblfloatfix,placeins,float,balance,needspace}
\hypersetup{hidelinks,pdfview=FitH,pdfstartview=FitH}

\renewcommand{\arraystretch}{0.95}
\title{Closed-Loop Dual-Reference Guidance for Accent Control in Cross-Lingual Voice Cloning}
\name{Yuanzhong Zheng, Wei Huang, Ming Ma, Kailun Jin, Youhai Jiang\sthanks{Corresponding author.}}
\address{VoxNexus, Singapore}

\begin{document}
\ninept
\maketitle
\begin{abstract}
Zero-shot cross-lingual voice cloning often carries the reference speaker's
accent into the target language. We propose closed-loop dual-reference
guidance, which pairs the original reference with a reusable native anchor
constructed by voice conversion (VC). The references initialize separate
language-model (LM) states that guide a shared acoustic trajectory; each
completed patch then updates both states, keeping their generated histories
aligned with the emitted speech. Anchor construction runs once per speaker,
and online synthesis uses a frozen TTS model without per-utterance VC. On 20
held-out French speakers, the fixed setting $\lambda^{\!*}=0.33$ reduces
residual French LID mass by 73\% versus direct cloning, at an ECAPA
similarity cost of only $0.042$; relative to anchor-only cloning, it recovers
$0.155$. The method achieves the lowest mean WER ($0.009$) and highest mean
GOP ($4.750$) among the held-out test arms. In blinded preference tests,
native-English listeners favor it over direct cloning on pronunciation
naturalness, and over both anchor-only cloning and a TTS$\to$VC cascade on
speaker identity. Online RTF is approximately 26\% lower than the cascade's
($0.59$ vs.\ $0.80$), and the same setting reduces Spanish LID mass by 56\%
without re-tuning.
\end{abstract}
\section{Introduction}
\label{sec:intro}

Cross-lingual zero-shot text-to-speech (TTS) synthesizes a target language
from a short reference in another language. The reference carries both vocal
identity and pronunciation patterns. A French reference, for example, can
yield intelligible English with a French accent. Accent control must therefore
regulate how reference pronunciation influences synthesis while retaining
speaker identity.

Accent control can be learned during training or applied during synthesis.
Disentanglement methods use multilingual or multi-accented data and
task-specific
optimization~\cite{xu2026xvoice,liu2023dsetts,cho2022sanetts,annamdevula2026crossaccent};
a TTS$\to$voice-conversion (VC) cascade instead converts every output
utterance. For frozen models, Activation Steering~\cite{yang2026steering} uses
offline-extracted accent directions, while standard classifier-free guidance
(CFG) and its flow-matching
adaptations~\cite{ho2021cfg,lipman2023flow,zheng2023guided} mix conditional and
unconditional velocity fields. We use a native speech reference to supply the
pronunciation condition and reuse it across utterances, placing accent control
inside autoregressive generation.

Our method uses two references with distinct, complementary roles: the
original speech anchors speaker identity, while a converted native anchor
supplies target-language pronunciation. Both references condition the same generated
patch, but mixing their decoder outputs alone leaves open how the two
autoregressive states should advance. We feed the emitted patch back to both
states, so each branch follows the speech actually produced while retaining
its own reference context. The weight $\lambda$ controls the anchor's
contribution, and the same anchor is reused for subsequent utterances.

\noindent\textbf{Contributions:}
\begin{enumerate}
\item \textbf{Closed-loop autoregressive composition.} We address context
desynchronization: independently advanced reference branches condition
guidance on different generated histories. Feeding each emitted patch back to
both LM states preserves a common history while retaining distinct reference
contexts. On development, this update reduces WER from $0.139$ with
independent branch rollouts to $0.004$ and raises GOP from $3.524$ to $4.791$.
\item \textbf{Reusable native anchors.} Zero-shot VC constructs a
speaker-specific native reference at enrollment, without parallel bilingual
recordings or task-specific training. Subsequent utterances reuse this
reference, moving conversion out of online synthesis.
\item \textbf{Composition in velocity and conditioning space.} The shared
state update supports both CFG-Zero*~\cite{fan2025cfgzero} velocity mixing and
fusion of decoder conditions. Conditioning fusion uses one conditional branch
instead of two, reducing decoder branch evaluations by one third ($27$ to $18$
per patch) while retaining both reference contexts and their feedback.
\end{enumerate}

\begin{figure*}[t]
  \centering
  \includegraphics[width=0.96\linewidth]{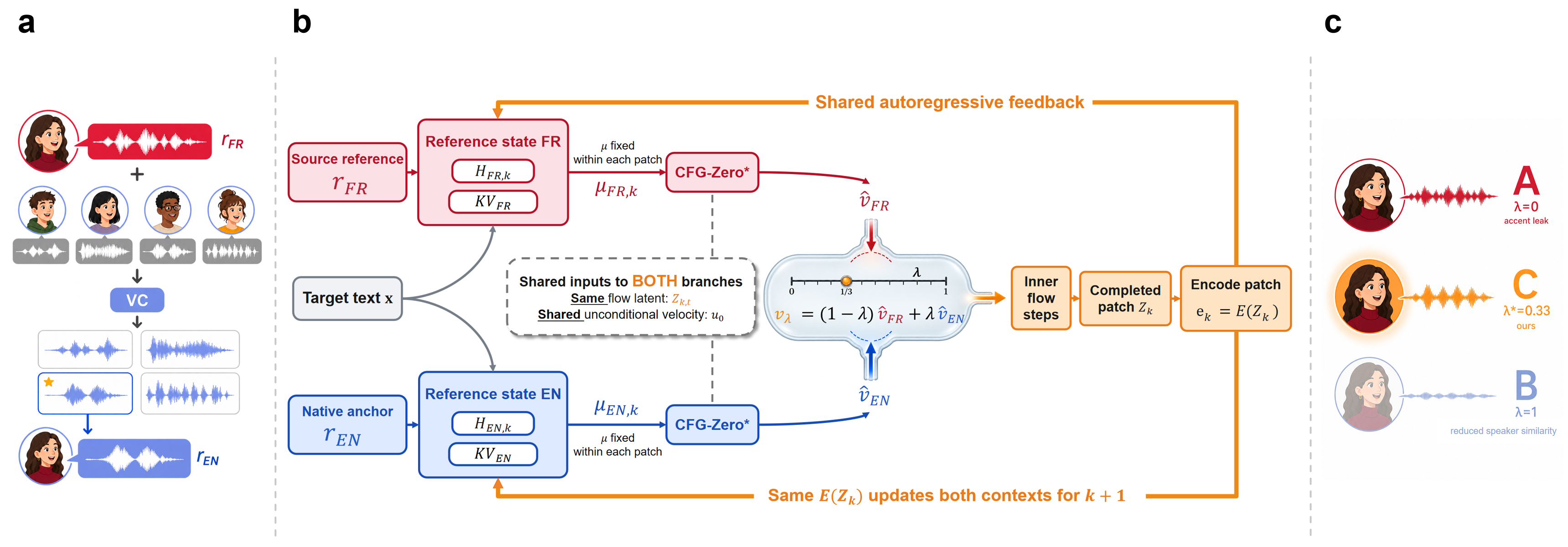}
  \caption{Method overview.
  \textbf{(a)}~Offline native-anchor construction: $r_{\mathrm{FR}}$ and four
  native donors undergo zero-shot conversion once per speaker; the
  highest-similarity candidate is kept as $r_{\mathrm{EN}}$.
  \textbf{(b)}~Closed-loop dual-reference guidance.
  At autoregressive step $k$, both reference-conditioned fields
  are evaluated on the same flow latent $z_{k,t}$ and combined via Eq.~(1).
  After the ODE produces acoustic patch $Z_k$, its encoding advances both LM
  contexts via Eq.~(2). Shared generated history; distinct
  reference contexts and KV caches.
  \textbf{(c)}~$\lambda=0$ is Arm~A (accent leakage); $\lambda=1$ is Arm~B
  (reduced speaker similarity); $\lambda^{\!*}=0.33$ is the selected operating
  point (Arm~C).}
  \label{fig:method}
\end{figure*}

\section{Related Work}
\label{sec:related}

\textbf{Accent conversion and post-hoc steering.}
Earlier work maps non-native speech toward a native
target~\cite{felps2009foreign,aryal2014can}, uses phonetic
posteriorgrams~\cite{zhao2019foreign,quamer2022zeroshot}, or factorizes
accent-related attributes in a neural codec~\cite{halychanskyi2026fac}.
Activation Steering~\cite{yang2026steering} extracts layer-wise accent
directions without further training. Our guidance comes from a
speaker-specific native anchor and two reference contexts updated with the
generated speech. Vevo~\cite{zhang2025vevo} separates style and timbre
through trained modules. We use Seed-VC~\cite{liu2024seedvc} for both anchor
construction and the TTS$\to$VC baseline, comparing a reusable enrollment-time
reference with per-utterance conversion.

\textbf{Cross-lingual TTS and guidance formulations.}
Zhang et al.~\cite{zhang2019learning} studied accent leakage in cross-lingual
TTS, followed by training-time disentanglement
methods~\cite{liu2023dsetts,cho2022sanetts,annamdevula2026crossaccent,lertpetchpun2026accent}.
X-Voice~\cite{xu2026xvoice} combines pseudo-parallel prompts for training-time
adaptation with dropped-condition CFG,
$v(\mathrm{ref})-v(\emptyset)$. CrossAccent-TTS~\cite{annamdevula2026crossaccent}
and Accent Vector~\cite{lertpetchpun2026accent} learn representations or
weights for accent control. Our method composes two reference-conditioned
states during inference in a frozen TTS model. Dual-scale CFG methods vary
guidance between text and speaker
conditions~\cite{yang2024dualspeech,jiang2025megatts,zheng2025selective}; our
two contexts each contain a speech reference and follow a common generated
history. Section~6.1 compares native-anchor guidance with a reference-free
secondary context.

\nocite{misra2014spoken,xia2019cross,estevez2023fairness,deja2022automatic,carbonneau2025analyzing}

\section{Method}
\label{sec:method}

\subsection{Synthetic native anchor}
\label{ssec:anchor}

Let $x$ denote target-language text, and let a frozen zero-shot TTS condition
its flow-matching decoder on reference speech $r$, yielding
$v(z,t;x,r)$. For each speaker we synthesize a native-English anchor
$r_{\mathrm{EN}}$ with a zero-shot voice converter
(Seed-VC)~\cite{liu2024seedvc}, converting a native English donor toward the
timbre of the source reference $r_{\mathrm{FR}}$.

Donors are a curated LibriTTS-R~\cite{koizumi2023librittsr} subset
(highest-DNSMOS utterance from each of the top 20 native English speakers).
With a fixed seed, four distinct donors are cyclically assigned per speaker.
Seed-VC with length adjustment $1.0$ preserves donor duration
($3.04$--$9.92$\,s). We keep the candidate maximizing
$\mathrm{SIM}_{\mathrm{avg}}=(\mathrm{SIM}_{\mathrm{ECAPA}}+\mathrm{SIM}_{\mathrm{CAM++\text{-}CN}})/2$
against $r_{\mathrm{FR}}$. The four conversions are performed at enrollment;
the selected anchor is reused for subsequent utterances (Fig.~1).

\subsection{Closed-loop dual-reference guidance}
\label{ssec:guidance}

Generation has an outer autoregressive patch index $k$ and an inner flow time
$t$. Let $H_{X,k}$ denote the base- and residual-LM state, including their KV
caches, for reference context $X\in\{\mathrm{FR},\mathrm{EN}\}$. Each context
is initialized by jointly prefilling the target text and local-encoder features
of $r_{X}$. At step $k$, the LMs provide diffusion-transformer (DiT)
conditioning $\mu_{X,k}$, which remains fixed during that patch's ODE solve.
Both fields are evaluated on the same evolving latent $z_{k,t}$ and combined
with CFG-Zero*~\cite{fan2025cfgzero}. Suppressing $k$ for clarity, define
$u_{X}=v(z,t;x,\mu_{X})$ and the shared unconditional velocity
$u_{0}=v(z,t;x,\emptyset)$. With stabilizer $\epsilon=10^{-8}$ and guidance
scale $s=2.0$,
\begin{equation}
\left\{
\begin{aligned}
\alpha_{X}
  &= \frac{\langle u_{X}, u_{0}\rangle}{\lVert u_{0}\rVert_{2}^{2}+\epsilon}, \\
\widehat{v}_{X}
  &= \alpha_{X} u_{0} + s\bigl(u_{X}-\alpha_{X} u_{0}\bigr), \\
v_{\lambda}
  &= (1-\lambda)\,\widehat{v}_{\mathrm{FR}} + \lambda\,\widehat{v}_{\mathrm{EN}}.
\end{aligned}
\right.
\label{eq:mix}
\end{equation}
CFG-Zero* splits the conditional field into components parallel and orthogonal
to $u_{0}$ via $\alpha_{X}$. The first Euler update is zeroed (fraction
$0.04$); the remaining nine follow Eq.~(1). Sampling is fixed (sway
$1.0$, temperature $1.0$, 10 Euler steps).

After the ODE produces acoustic patch $Z_k$, its local encoding and projection
$E(Z_k)$ advance both LM contexts:
\begin{equation}
H_{X,k+1}=F\bigl(H_{X,k},E(Z_k)\bigr),
\qquad X\in\{\mathrm{FR},\mathrm{EN}\}.
\label{eq:feedback}
\end{equation}
Here $F$ is the frozen base- and residual-LM update. Both contexts follow the
emitted history while retaining distinct reference prefills. The open-KV
ablation uses independently sampled histories (Sec.~6.1).

The same update supports a conditioning-space implementation:
$\mu\leftarrow(1-\lambda)\mu_{\mathrm{FR}}+\lambda\mu_{\mathrm{EN}}$. This
variant retains both contexts and Eq.~(2), then uses a single conditional DiT
branch plus the unconditional branch (Secs.~5.3, 6.1). Velocity mixing is used
for all held-out test and transfer runs. $\lambda=0$ gives A, $\lambda=1$
gives B, and intermediate values give~C. We swept
$\lambda\in\{0.1,0.2,0.25,0.33,0.5,0.75\}$ on development data and fixed
$\lambda^{\!*}=0.33$ before evaluating the test set; throughout,
$\lambda^{\!*}$ denotes this selected operating point rather than an
optimized value.

\section{Experimental Setup}
\label{sec:expt}

\begin{table*}[t]
\centering
\caption{Main evaluation on held-out test data (20 French speakers $\times$ 20
texts $\times$ 3 seeds; $n=1200$ per arm, $\lambda^{\!*}=0.33$). Speaker-macro
averages; bold values mark the best result in each column. $P(\mathrm{fr})$ is
residual French LID mass. $^{*}$CAM++-CN is Seed-VC's internal speaker encoder
(Sec.~6.2).}
\label{tab:main}
\footnotesize
\setlength{\tabcolsep}{3.0pt}
\begin{tabular}{lrrrcccrr}
\hline
 & WER $\downarrow$ & $P(\mathrm{fr})$ $\downarrow$ & $P(\mathrm{en})$ $\uparrow$ &
 \multicolumn{3}{c}{Speaker similarity $\uparrow$} & GOP $\uparrow$ & DNSMOS $\uparrow$ \\
\cline{5-7}
Arm &  &  &  & ECAPA & CAM++-EN & CAM++-CN$^{*}$ &  &  \\
\hline
A (Direct) & 0.016 & 0.093 & 0.843 & \textbf{0.601} & \textbf{0.624} & 0.746 & 4.663 & 2.900 \\
B (Anchor) & 0.013 & \textbf{0.000} & \textbf{0.993} & 0.404 & 0.444 & 0.702 & 4.733 & 2.934 \\
\textbf{C (Proposed)} & \textbf{0.009} & 0.025 & 0.942 & 0.559 & 0.586 & 0.748 & \textbf{4.750} &
2.887 \\
Cascade & 0.047 & 0.001 & 0.985 & 0.468 & 0.492 & \textbf{0.770} & 4.378 & \textbf{3.026} \\
\hline
\end{tabular}
\end{table*}

\textbf{Models.}
Frozen VoxCPM2~\cite{zhou2026voxcpm2} (2B, 48\,kHz, tokenizer-free, no LoRA).
Anchors and the cascade use Seed-VC~\cite{liu2024seedvc} (Whisper-small $+$
BigVGAN-v2; 25 steps, CFG $0.7$, length adjustment $1.0$).

\textbf{Arms.}
We refer to each compared configuration as an arm. A: real $r_{\mathrm{FR}}$.
B: synthetic $r_{\mathrm{EN}}$. C: proposed guidance at $\lambda^{\!*}=0.33$.
Cascade: unconditioned English from VoxCPM2, then Seed-VC transfers
$r_{\mathrm{FR}}$ timbre on every utterance.

\textbf{Data.}
Development (VP-Stress): 30 VoxPopuli~\cite{wang2021voxpopuli} French speakers
(LID-FR $\geq 0.9$, reference $3$--$10$\,s), 10 English texts, seed 20260822
($n=300$/arm). Held-out test (B1-CNS-test): 20 MLS~\cite{pratap2020mls} French
speakers $\times$ 20 English texts $\times$ 3 seeds, item-matched $n=1200$/arm
including the cascade; speakers are disjoint from development, donors, and
tuning. All 20 test anchors pass $\mathrm{SIM}_{\mathrm{avg}}>0.45$ (mean
$0.707$). Generalization: 15 VoxPopuli Spanish speakers, 10 English texts, with
$\lambda^{\!*}=0.33$ transferred without re-tuning.

\textbf{Metrics.}
WER and marginal $P(\mathrm{fr})$ / $P(\mathrm{es})$:
faster-whisper-large-v3~\cite{radford2023whisper} (beam 1, no VAD; full softmax,
not top-1). GOP~\cite{witt2000phone}:
wav2vec2-xls-r-300m-phoneme~\cite{babu2022xlsr} posteriors, g2p\_en, CTC
alignment; utterance GOP averages $\log P(y)-\max_{p\neq y}\log P(p)$ over
target phones $y$. Similarity: cosine vs.\ the foreign reference on
ECAPA-TDNN~\cite{desplanques2020ecapa}, CAM++-EN, and
CAM++-CN~\cite{wang2023campplus}. We report all three encoders and listener
preferences to cover language/accent
sensitivity~\cite{misra2014spoken,xia2019cross,estevez2023fairness} and
differences between automatic and perceptual similarity
judgments~\cite{deja2022automatic,carbonneau2025analyzing}. Quality: DNSMOS
P.835 OVRL~\cite{reddy2022dnsmos}. Statistics: speaker-macro means, paired
deltas, 10{,}000 speaker-clustered bootstrap 95\% CIs; preferences use session
majority and a speaker-clustered two-sided exact sign-flip test.

\section{Results}
\label{sec:results}

\subsection{Development trade-off and \texorpdfstring{$\lambda^{*}$}{lambda*} selection}
\label{ssec:tradeoff}

At $\lambda=0.33$, the GOP gain over A is $+0.189$, nearly matching the
$+0.193$ attained at $\lambda=0.5$, while the ECAPA similarity cost is
roughly halved ($\Delta\mathrm{ECAPA}=-0.049$ [$-0.063,-0.036$]; Fig.~2).
CAM++-EN follows the same trend. At this setting, paired differences are $+0.021$ [$-0.062,0.098$] in
GOP relative to B and $-0.018$ [$-0.068,0.032$] in DNSMOS relative to the
cascade. We selected $\lambda^{\!*}=0.33$ on development and fixed it for test
and Spanish transfer.

\begin{figure}[!ht]
  \centering
  \includegraphics[width=0.62\linewidth]{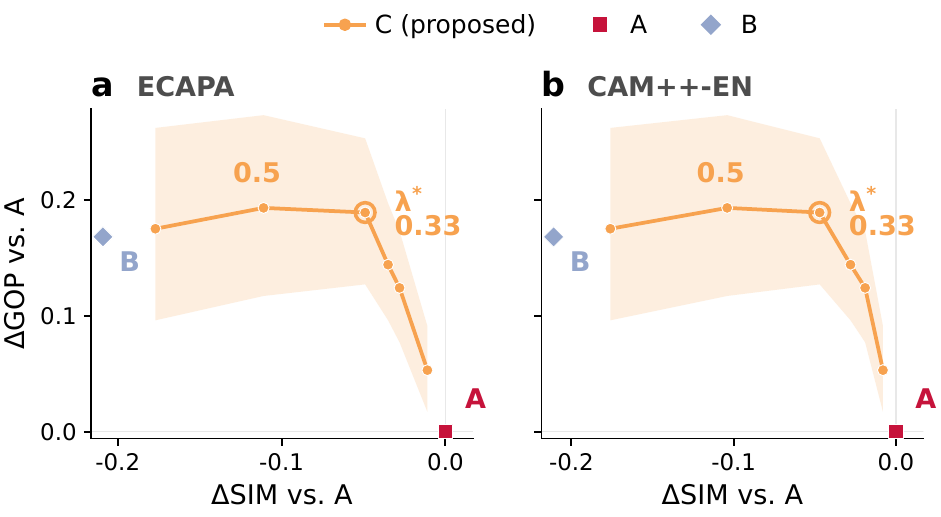}
  \caption{Accent--timbre trade-off on development (30 speakers).
  \textbf{(a)}~ECAPA; \textbf{(b)}~CAM++-EN.
  $x$: $\Delta\mathrm{SIM}$ vs.\ A; $y$: $\Delta\mathrm{GOP}$ vs.\ A. The
  shaded band is the 95\% speaker-clustered bootstrap CI of $\Delta\mathrm{GOP}$.
  Orange: C (proposed) $\lambda$ sweep; $\lambda^{\!*}$ is the selected
  operating point ($0.33$).}
  \label{fig:tradeoff}
\end{figure}

\subsection{Held-out test evaluation (\texorpdfstring{$\lambda^{*}$}{lambda*} frozen)}
\label{ssec:main}

\textbf{C vs.\ A.}
At the fixed operating point, C achieves mean WER $0.009$ and GOP $4.750$, the
lowest and highest values, respectively, among the held-out test arms
(Table~1). Relative to direct cloning, it reduces residual French LID mass by
73\% ($0.093\to 0.025$; $\Delta=-0.068$ [$-0.119,-0.027$]) and increases GOP
by $0.087$ [$0.048,0.128$] (Fig.~3). The paired WER difference is $-0.0072$
[$-0.0147,0.0003$], with an ECAPA change of $-0.042$ [$-0.048,-0.036$].

\textbf{C vs.\ B.}
Adding the original-reference branch recovers $0.155$ ECAPA similarity
([$0.137,0.173$]) and $0.142$ CAM++-EN similarity ([$0.122,0.162$]) over
anchor-only cloning. This speaker-similarity recovery comes with higher mean
GOP ($4.750$ vs.\ $4.733$) and lower mean WER ($0.009$ vs.\ $0.013$); paired
differences are $+0.017$ [$-0.043,0.072$] and $-0.004$ [$-0.011,0.002$].

\begin{figure}[!ht]
  \centering
  \includegraphics[width=0.78\linewidth]{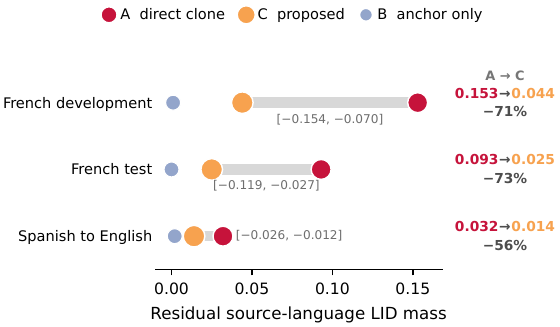}
  \caption{Residual source-language LID mass at $\lambda^{\!*}=0.33$.
  Dumbbells: A$\to$C reduction; B is the floor. Brackets give
  speaker-clustered 95\% CIs of the paired difference $\Delta=C{-}A$
  (negative values indicate a reduction under C).}
  \label{fig:lid}
\end{figure}

\subsection{Comparison with the TTS\texorpdfstring{$\to$}{->}VC cascade}
\label{ssec:cascade}

On the item-matched three-seed test set ($n=1200$/arm), C reduces mean WER
from the cascade's $0.047$ to $0.009$, an approximately 81\% reduction
($\Delta=-0.038$ [$-0.046,-0.031$]), and raises GOP from $4.378$ to $4.750$
($\Delta=+0.372$ [$0.330,0.412$]). Word repetition or insertion involving at
least two words occurs in $3.2\%$ ($38/1200$) of cascade outputs, versus less
than $0.5\%$ for A, B, and C.

Speaker similarity also improves over the cascade: ECAPA increases by $0.091$
(95\% CI [$0.073,0.109$]) and CAM++-EN by $0.094$ ([$0.073,0.114$]). In the
blinded development study, listeners likewise prefer C on speaker identity
($38$ vs.\ $5$ sessions, $p<0.0001$; Sec.~5.5). On the remaining metrics, C
trails the cascade: C$-$cascade is $+0.024$ on residual $P(\mathrm{fr})$
([$0.005,0.052$]; cascade mean $0.001$), $-0.139$ on DNSMOS
([$-0.197,-0.089$]), and $-0.022$ on CAM++-CN (CI [$-0.039,-0.006$];
Sec.~6.2). Thus, the cascade's higher DNSMOS accompanies more word errors,
while C improves both transcription accuracy and listener-rated speaker
identity.

\begin{table}[!ht]
\centering
\caption{H100 80GB inference. \textbf{(a)}~vs.\ direct cloning and cascade.
\textbf{(b)}~Matched velocity vs.\ conditioning ($\mu$-mix). TTS: 9-word text,
five timed runs after warmup. DiT: batched branch evaluations per patch over
nine nonzero Euler updates. $^{\dagger}$sequential cascade stages.}
\label{tab:compute}
\footnotesize
\setlength{\tabcolsep}{2.0pt}
\begin{tabular}{lccccc}
\hline
Arm & DiT/patch & VC NFE & Lat.\ (s) & RTF & Peak VRAM (MiB) \\
\hline
\multicolumn{6}{l}{\textit{(a) Direct cloning and cascade comparison}} \\
A & 18 & --- & 1.27 & 0.474 & 8239 \\
\textbf{C (velocity)} & 27 & --- & 1.52 & 0.587 & 8235 \\
Cascade & 18 & 50 & $\approx$2.10 & $\approx$0.80 & 8563 / 1474$^{\dagger}$ \\
\hline
\multicolumn{6}{l}{\textit{(b) Matched implementation comparison}} \\
C (velocity) & 27 & --- & 1.62 & 0.589 & 8406 \\
C (conditioning) & 18 & --- & 1.47 & 0.572 & 8408 \\
\hline
\end{tabular}
\end{table}

\textbf{Compute.}
C (velocity) achieves approximately 26\% lower online RTF than the cascade
($0.59$ vs.\ $0.80$). Relative to direct cloning, it adds $0.25$\,s with
essentially unchanged peak memory. Four anchor conversions cost approximately
$3.1$\,s per speaker. At the benchmarked length, $T_{C}(N)\approx 3.1+1.52N$
and $T_{\mathrm{cascade}}(N)\approx 2.10N$; the reusable anchor amortizes
conversion cost by six utterances.

\textbf{Conditioning-space implementation.}
C (conditioning) reduces DiT branch evaluations from 27 to 18 per patch and
retains both LM contexts. In the matched rerun, its RTF is $0.572$ versus
$0.589$ for C (velocity), with nearly identical peak memory. RTF reductions
across 5-, 9-, and 20-word texts are $0.7$--$2.9\%$.

\subsection{Cross-language generalization (Spanish\texorpdfstring{$\to$}{->}English)}
\label{ssec:spanish}
\balance

The French-selected $\lambda^{\!*}=0.33$ transfers directly to 15 Spanish
speakers without language-specific tuning. Residual Spanish LID mass drops
from $0.032$ to $0.014$ ($\Delta=-0.018$ [$-0.026,-0.012$]; Fig.~3), and
ECAPA similarity recovers from $0.323$ to $0.474$ relative to B. C achieves
WER $0.0055$, compared with $0.0054$ for A
($\Delta=+0.0001$ [$-0.0070,0.0065$]). The same operating point therefore
combines source-LID reduction with speaker-similarity recovery in both
French-to-English and Spanish-to-English cloning.

\subsection{Subjective human evaluation}
\label{ssec:listening}

A randomized, blinded preference study used seven native-English evaluators and
60 sessions (30 development speakers $\times$ 2 texts), with the French
reference and outputs from A, B, C ($\lambda^{\!*}=0.33$), and the cascade.
Evaluators chose the baseline, C, or a tie. Four pre-planned comparisons
covered accent (A vs.\ C, C vs.\ cascade) and speaker similarity (B vs.\ C,
C vs.\ cascade). A follow-up comparison assessed C vs.\ A speaker similarity.

\begin{table}[H]
\centering
\caption{Preferences on development speakers (7 evaluators $\times$ 60 sessions).
\% (sessions), majority vote. Bold: significantly preferred system. $p$:
speaker-clustered sign-flip; pre-planned rows remain significant under
Holm--Bonferroni. $^{\dagger}$Follow-up speaker-similarity comparison.}
\label{tab:pref}
\footnotesize
\setlength{\tabcolsep}{2.0pt}
\begin{tabular}{llrrrc}
\hline
Comp. & Dim. & Pref.\ base & Pref.\ C & Tie & $p$ \\
\hline
C vs.\ A & Accent / nat. & 6.7 (4) & \textbf{28.3 (17)} & 65.0 (39) &
\textbf{$0.0057$} \\
C vs.\ B & Speaker sim. & 5.0 (3) & \textbf{65.0 (39)} & 30.0 (18) &
\textbf{$<0.0001$} \\
C vs.\ Casc. & Accent / nat. & \textbf{53.3 (32)} & 21.7 (13) & 25.0 (15) &
\textbf{$0.026$} \\
C vs.\ Casc. & Speaker sim. & 8.3 (5) & \textbf{63.3 (38)} & 28.3 (17) &
\textbf{$<0.0001$} \\
C vs.\ A$^{\dagger}$ & Speaker sim. & 26.7 (16) & 21.7 (13) & 51.7 (31) &
$0.723$ \\
\hline
\end{tabular}
\end{table}

Listeners significantly prefer C to direct cloning on pronunciation
naturalness ($17$ vs.\ $4$ sessions, $p=0.0057$; Table~3). In the follow-up
speaker-similarity comparison with direct cloning, $51.7\%$ of sessions are
ties; the remaining sessions split $13$ for C and $16$ for A ($p=0.723$).
Relative to anchor-only cloning, C wins $39$ vs.\ $3$ sessions on speaker
identity ($p<0.0001$). The comparison with the cascade separates pronunciation
from identity: listeners favor the cascade on pronunciation naturalness
($32$ vs.\ $13$, $p=0.026$), and C on speaker identity ($38$ vs.\ $5$,
$p<0.0001$). Fleiss' $\kappa$ is $0.167$--$0.256$; 87--89\% of
majority-disagreement judgments are ties.

\Needspace{8\baselineskip}
\section{Analysis}
\label{sec:analysis}

\subsection{Guidance mechanisms and architectural ablations}
\label{ssec:ablate}
\nopagebreak
\textbf{Shared autoregressive feedback.}
The open-KV ablation tests the effect of independently advancing the reference
branches. Both configurations mix velocities for the emitted patch, but
open-KV feeds each LM its own separately sampled, unmixed prediction. This
permits context desynchronization: each branch follows its own generated
history rather than the emitted sequence. Replacing these independent rollouts
with shared patch feedback reduces WER from $0.139$ to $0.004$ and raises GOP
from $3.524$ to $4.791$. Equation~(2) enforces the common-history constraint
at every patch while preserving each branch's reference context.

\textbf{Reference contexts.}
C (velocity) gives ECAPA $0.467$ versus $0.372$ for concat; concat reaches
$P(\mathrm{fr})=0.003$. Replacing the native anchor with $\emptyset$ lowers
similarity and GOP. The waveforms used by wavavg differ in length and phase;
their average gives GOP $3.131$.

\textbf{Composition across mixing sites.}
C (conditioning) achieves ECAPA $0.465$ and GOP $4.828$, versus $0.467$ and
$4.791$ for C (velocity). The paired differences are not significant
(velocity minus conditioning: ECAPA $+0.003$ [$-0.004,0.009$]; GOP $-0.038$
[$-0.089,0.014$]). Both implementations retain separate reference states and
common patch feedback; conditioning mixing achieves RTF $0.572$ (Table~2).

\begin{table}[H]
\centering
\caption{Paired ablations on development (30 speakers, $\lambda=0.33$,
$n=300$).}
\label{tab:ablate}
\footnotesize
\setlength{\tabcolsep}{2.4pt}
\renewcommand{\arraystretch}{0.90}
\begin{tabular}{lrrrrr}
\hline
Config. & ECAPA $\uparrow$ & CAM++-EN $\uparrow$ & GOP $\uparrow$ & WER $\downarrow$
& $P(\mathrm{fr})$ $\downarrow$ \\
\hline
A ($\lambda=0$) & 0.517 & 0.589 & 4.602 & 0.020 & 0.153 \\
B (anchor only) & 0.308 & 0.378 & 4.770 & 0.009 & 0.001 \\
\textbf{C (velocity)} & 0.467 & 0.542 & 4.791 & 0.004 & 0.044 \\
C (conditioning) & 0.465 & 0.546 & 4.828 & 0.000 & 0.034 \\
concat & 0.372 & 0.459 & 4.743 & 0.003 & 0.003 \\
open-KV & 0.458 & 0.522 & 3.524 & 0.139 & 0.045 \\
$C_{\mathrm{noref}}$ ($\emptyset$) & 0.441 & 0.517 & 4.653 & 0.010 & 0.063 \\
wavavg & 0.359 & 0.420 & 3.131 & 0.011 & 0.008 \\
\hline
\end{tabular}
\end{table}

\subsection{Encoder-domain discrepancy in speaker verification}
\label{ssec:simenc}

The three encoders assign substantially different similarity changes to the
same accent-guided speech. On French development, C vs.\ A at $\lambda=0.5$ is
$-0.111$ on ECAPA and $-0.104$ on CAM++-EN, but $-0.020$ on CAM++-CN. The
CAM++ checkpoints share an architecture but were trained on VoxCeleb2 and a
200k-speaker Chinese dataset, respectively; their responses show that the
estimated identity cost of accent control depends on the verification model.
At $\lambda^{\!*}=0.33$, the CAM++-CN difference is $+0.003$
[$-0.003,0.008$]. CAM++-CN also serves as Seed-VC's internal speaker encoder
and contributes to our anchor-selection score (Sec.~3.1). On the held-out
test, ECAPA and CAM++-EN rank C above the cascade, whereas CAM++-CN reverses
that ordering. The blinded listening study independently favors C on speaker
identity ($38$ vs.\ $5$ sessions; Sec.~5.5), aligning with the ECAPA and
CAM++-EN ranking.

\section{Conclusion and Limitations}
\label{sec:conclusion}
We presented closed-loop dual-reference guidance for accent control in frozen
TTS. Separate reference states combine original-speaker information with a
reusable native anchor, and feedback from the emitted patches keeps both
branches on the same generated history; this common-history constraint,
rather than the choice of mixing site, accounts for the largest quality gain
(Sec.~6.1). On held-out French speakers, the method cuts residual French LID
mass by 73\% versus direct cloning while recovering $0.155$ ECAPA similarity
over anchor-only cloning, and the French-selected operating point transfers
to Spanish without re-tuning. Native-English listeners prefer C on
pronunciation naturalness over A and on speaker identity over B and the
cascade. The current evaluation targets English only, anchor quality depends
on the zero-shot converter, and the guidance weight is fixed across patches;
extending the evaluation to additional target languages, adapting guidance
per patch, and improving converted-anchor quality are the next steps.

\clearpage
{\footnotesize
\linespread{0.84}\selectfont
\setlength{\itemsep}{-4.0pt plus 0.05pt}
\setlength{\parsep}{0pt}
\setlength{\parskip}{0pt}
\bibliographystyle{IEEEbib}
\bibliography{refs}
}

\end{document}